# Memory Effect and Triplet Pairing Generation in the Superconducting Exchange Biased Co/CoO$_x$/Cu$_{41}$Ni$_{59}$/Nb/Cu$_{41}$Ni$_{59}$ Layered Heterostructure

V.I. Zdravkov[1,2], D. Lenk[1], R. Morari[1], A. Ullrich[1], G. Obermeier[1], C. Müller[1],
H.-A. Krug von Nidda[1], A.S. Sidorenko[2], S. Horn[1], R. Tidecks[1], and L.R. Tagirov[1,3]

[1]*Institut für Physik, Universität Augsburg, D-86159 Augsburg, Germany*
[2]*D. Ghitsu Institute of Electronic Engineering and Nanotechnologies ASM, MD2028 Kishinev, Moldova*
[3]*Solid State Physics Department, Kazan Federal University, 420008 Kazan, Russia*

We fabricated a nanolayered hybrid superconductor-ferromagnet spin-valve structure, the resistive state of which depends on the preceding magnetic field polarity. The effect is based on a strong exchange bias (about -2 kOe) on a diluted ferromagnetic copper-nickel alloy and generation of a long range odd in frequency triplet pairing component. The difference of high and low resistance states at zero magnetic field is 90% of the normal state resistance for a transport current of 250 µA and still around 42% for 10 µA. Both logic states of the structure do not require biasing fields or currents in the idle mode.

Superimposing two antagonistic phenomena, superconductivity (S) and ferromagnetism (F), on the nanoscale offers rich basic physics[1-3] and provides several opportunities to design superconducting devices having unique features.[4-7] The S/F interaction can in general be described in terms of the proximity effect with a mutual penetration of charge carriers, electrons or Cooper pairs, or stray field mediated correlations. Superconducting spin-valves (SSVs)[8-10], intended to switch between two states with different superconducting transition temperatures, $T_c$, show extremely large magnetoresistance and are under wide experimental and theoretical consideration since the last decade.[11-18] In both, F/S/F and S/F/F designs of the SSV, $T_c$ is manipulated by altering the magnetic configuration of the F-layers. A positive difference in $T_c$ between anti-parallel (AP) and parallel (P) configuration, $\Delta T_c^{AP-P}$, is described in terms of the S/F proximity effect,[8-16] while stray-field mediated mutual correlations of micromagnetic structures in the F-layers is a plausible explanation of the negative $\Delta T_c^{AP-P}$ in F/S/F spin-valves[17-19].

Unconventional odd-triplet pairing, recently considered for S/F proximity systems,[2] deepens the understanding and extends the functionality of the superconducting spin–valves,[20-23] introducing a coupling by long-range spin-polarized Cooper pairs. As a result, $\Delta T_c^{AP-P}$ in S/F/F-type SSVs can be either positive or negative within the proximity coupling model. Moreover, $T_c$ can have an absolute minimum at non-collinear alignments of the F-layer magnetic moments, resulting in the triplet switching mode.[20]

Control of magnetic configurations in spin-valves is often provided by bringing one of the F-layers in contact with an antiferromagnetic (AF) layer. The interfacial exchange coupling induces a unidirectional magnetic anisotropy, the exchange bias effect, which gives rise to a horizontal shift of the hysteresis loop, coercivity enhancement, asymmetric hysteresis loops, and training effects[24-28].

The exchange bias phenomenon is widely explored in magnetic field sensors,[25-27] however, even now it is not thoroughly understood and hardly predictable for an arbitrary AF-F couple of materials. In particular, a realization of a spin-valve device making use of weak disordered ferromagnetic alloys still remains an unresolved problem (see, e.g. Ref. [28] and citations therein). In this Letter we report on a strong exchange biasing for the $Cu_{41}Ni_{59}$ layer adjacent to the $Co/CoO_x$ interface in a $Co/CoO_x/Cu_{41}Ni_{59}/Nb/Cu_{41}Ni_{59}$ SSV structure. The magnetoresistive switching properties obtained make the heterostructure suitable for superconducting spintronics applications.

A set of samples was produced in one run by magnetron sputtering, mainly at room temperature. The sketch of the resulting stack is shown in Fig. 1(a). First, a metallic Co layer was deposited on a commercial (111) silicon substrate, covered by a Si buffer layer before. Next, reactive oxygen gas was mixed to argon to deposit a $CoO_x$ oxide layer. Subsequently, a $Cu_{40}Ni_{60}$ target was RF sputtered at 200 °C at a rate of 3 nm/sec, resulting regularly in a $Cu_{41}Ni_{59}$ composition of the alloy in the film, as checked by the Rutherford backscattering spectrometry (RBS) technique and scanning Auger spectroscopy (see details in Refs. [29-31]). To get a set of samples with different thicknesses of the $Cu_{41}Ni_{59}$ layer, the wedge technique[29-31] was applied. Thus, we obtained copper-nickel layer thicknesses as follows: maximum for Sample I, minimum or vanishing for Sample IV, and intermediate for adjacent Samples II and III. A flat superconducting Nb layer was prepared applying the "spray" technique[29-31]. To control precisely the film growth rate we monotonously moved the target during the DC sputtering process along the substrate. Thus, we achieved an effective growth rate of about 1.3 nm/sec for the Nb film, while the rate of the sputtering process was adjusted to 4 nm/sec, to reduce contaminations gettered into the Nb film. In this way we obtained a smooth Nb film of constant thickness of $d_{Nb} \approx 12$ nm. Finally, the stack was finished by depositing a second wedge-shaped copper-nickel layer and capped with 12-14 nm of silicon to protect it against oxidation.

To obtain the thicknesses of the layers, we used cross-sectional Transmission Electron Microscopy (TEM) measurements. For Sample I (see Fig. 1(a)) the thicknesses were determined as about 4, 14, 25, 13, and 22 nm for Co, $CoO_x$, CuNi-Bottom, Nb, and CuNi-Top layers, respectively, whereas from the TEM image of Sample II we got about 5, 19, 9, 12, and 10 nm. We evaluated the thicknesses of the layers for Samples III and IV as 5, 19, 8, 12, 9 nm and 4, 14, ≤ 1, 11.5, ≤ 1 nm for Co, $CoO_x$, CuNi-Bottom, Nb, and CuNi-Top, respectively, by extrapolation, applying, in addition, our experience in evaluating the wedge profile[29-31].

To explore the magnetic configurations of the $Co/CoO_x/Cu_{41}Ni_{59}/Nb/Cu_{41}Ni_{59}$ system, several hysteresis loops were subsequently measured by a Superconducting Quantum Interference Device (SQUID) magnetometer after cooling the samples from above the CoO Néel temperature (291 K) to 10 K in a field of 10 kOe, applied parallel to the heterostructure plane. First, the applied magnetic field was swept from the saturated state achieved in the field cooling direction, towards negative fields until saturation of the layers in the opposite direction is reached ("backward branch" (BB) of the $m(H)$ hysteresis loops). Then, the magnetic field was swept from negative fields to positive ones until saturation of all ferromagnetic layers was achieved ("forward branch" (FB)). The resulting dependences of the magnetic moment on the magnetic field, $m(H)$, as well as their derivatives, $\partial m/\partial H$ (*i.e.*, magnetic susceptibilities), for Sample III are shown in Fig. 2. For the first cycle see Fig. 2(a), for the repeated cycle (the second hysteresis loop and its derivative) see Fig. 2(b).

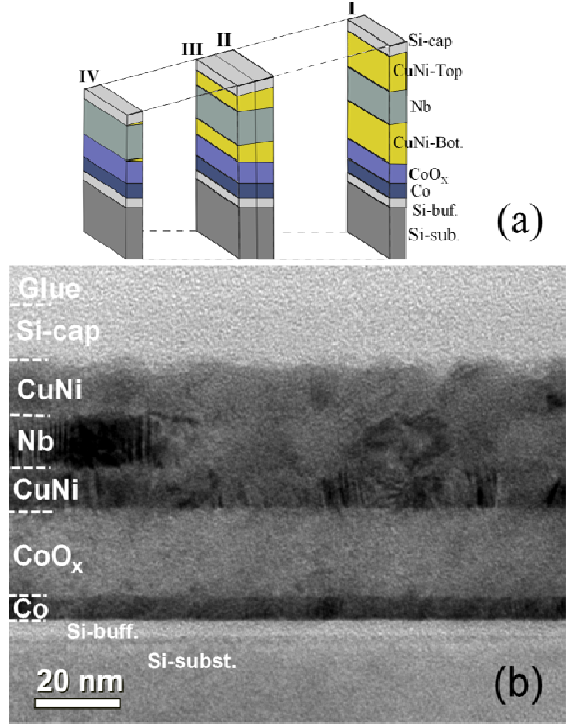

Fig. 1 (a) Sketch of the sample design; (b) TEM image of Sample II. In the present work Sample I and II are used for TEM images, while measurements of magnetic moment and resistance were performed on Samples III and IV under a magnetic field applied parallel to the layers. Dashed lines in (a) are guides to the eye and do not represent the real thickness gradient.

An apparent strong exchange bias is evident from the figure, as well as the training-effect[27], *i.e.* a decrease of the hysteresis loop asymmetry, coercivity and squareness by further magnetic field cyclings (Fig. 2(a) to Fig. 2(b)). The bottom $Cu_{41}Ni_{59}$ layer, adjacent to the $CoO_x$ layer, shows a very strong exchange bias $H_{EB} \approx$ -2 kOe, where $H_{EB} = (H_c^{BB} + H_c^{FB})/2$. It is clearly distinguishable from the peaks of the *m*(*H*) derivative labeled $H_c^2$ in Fig. 2(b), and in this magnitude never reported before for copper-nickel ferromagnetic alloys[11,28]. Particular magnetic configurations that can be imagined using the Stoner-Wohlfarth model (see, *e.g.*, Section 3.2 in Ref. [24]) of the magnetic moment reversal are depicted in Fig. 2 by the bold arrows. It could be concluded from Fig. 2(a), an antiparallel (AP) alignment of the magnetic moments of the top and bottom $Cu_{41}Ni_{59}$ layer, which is necessary to observe the direct spin-valve effect (yielding $\Delta T_c^{AP-P} > 0$, is achieved over the wide range, -1.2 to -4.0 kOe, of magnetic fields.

To explore the superconducting spin-valve effects we measured the magnetoresistance, *R*(*H*), at fixed temperatures in the range of the superconducting transition, which is most sensitive to the magnetic configurations in the system. The results are presented in Fig. 3. The standard DC four-terminal method was used, applying a sensing current of 10 µA. The polarity of the current was alternated during the resistance measurements to eliminate possible thermoelectric voltages.

The *R*(*H*) measurements start with a BB sweep from the positively saturated state (Fig. 3(a)), which is the starting point of the magnetic hysteresis measurements in Fig. 2(a). The top $Cu_{41}Ni_{59}$ layer reverses its magnetization and enters a single-domain state at about -1.2 kOe. The bottom $Cu_{41}Ni_{59}$ layer is expected to save its single domain state up to about -4 kOe due to strong exchange bias. Surprisingly, no indications of the expected *direct* superconducting spin-valve effect[9,10,29], $\Delta T_c^{AP-P} > 0$, was detected. Instead, a tiny *inverse* spin-valve effect (reduction of $T_c$ by less than 1 mK) can be deduced from the *R(H)* data in the range -1.2 to -4 kOe (see dashed line in Fig. 3(a)). Moreover, a clear abrupt resistance change is also visible upon the Co plus $Cu_{41}Ni_{59}$-bottom layers reversal at -4.4 kOe (compare Fig. 2(a) with Fig. 3(a)).

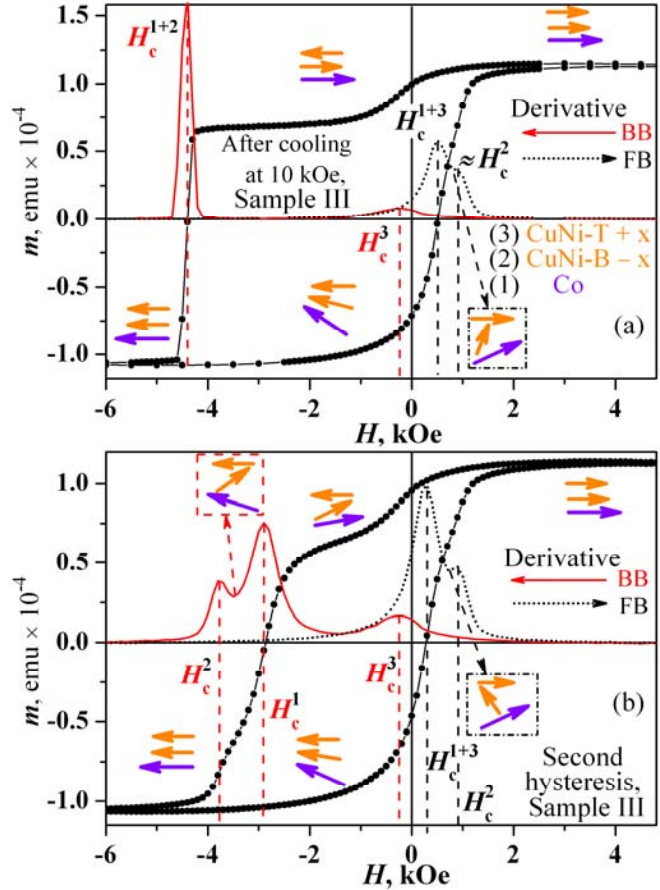

Fig. 2 (a) The magnetic hysteresis loop of a $Co/CoO_x/Cu_{41}Ni_{59}/Nb/Cu_{41}Ni_{59}$ nanolayered heterostructure (Sample III) measured in a magnetic field parallel to the layers just after cooling down in a magnetic field of 10 kOe. Thin arrows refer to the derivative of the magnetic moment against the field: red solid lines are for the BB, and black dotted lines for the FB. Bold arrows, drawn according the Stoner-Wohlfart model, indicate the orientation of the magnetization of layers (1) to (3), from bottom to top; (b) The second hysteresis loop measured after the previous one and showing the training effect. The derivatives clearly show a splitting of the reversal fields for the BB, and their partial overlapping for the FB. Here, *x* denotes a certain thickness of the bottom $Cu_{41}Ni_{59}$ layer (depending on *H* and the magnetic history, probably increasing by training effects), the magnetization of which follows the top $Cu_{41}Ni_{59}$ layer in the proposed model.

The absence of the standard (direct) superconducting spin-valve effect can be explained if no antiparallel alignment of the top and bottom $Cu_{41}Ni_{59}$ layers occurs. A seeming contradiction with the presence of the nearly flat region in the BB of the hysteresis in Fig 2(a), after the top $Cu_{41}Ni_{59}$ reversal, could be resolved if we assume the bottom $Cu_{41}Ni_{59}$ layer is an exchange spring[32,33]. A Nb side portion *x* of the thickness of the bottom $Cu_{41}Ni_{59}$ layer rotates its magnetization at almost the same field as the top, soft $Cu_{41}Ni_{59}$ layer does, while the exchange biased interface to $CoO_x$ and the rest of the layer keeps the initial direction. There is a region of gradual transition between the oppositely magnetized sub-layers, which can be treated as an exchange spring or a domain wall, depending on its extent. One can expect that the non-uniform distribution of the magnetic moments in this case may generate triplet pairing components[2,34-36], which suppress superconductivity acting hereby against the expected direct spin-valve effect[9,10]. A decomposition of the net hysteresis loop of Fig. 2(b) on its components shows that the response of the soft $Cu_{41}Ni_{59}$ magnetic moment is ~1.7 times larger than the response of the hard $Cu_{41}Ni_{59}$ magnetic moment, whereas from the ratio of the $Cu_{41}Ni_{59}$ layers thicknesses a value of ~1.13 is expected (9 nm for the top CuNi layer against 8 nm for the bottom one, see Fig. 1(b)). Thus, this observation favors the spring-magnet interpretation. The signal from the Co layer can be confidently separated from the above signals ascribed to the $Cu_{41}Ni_{59}$ material.

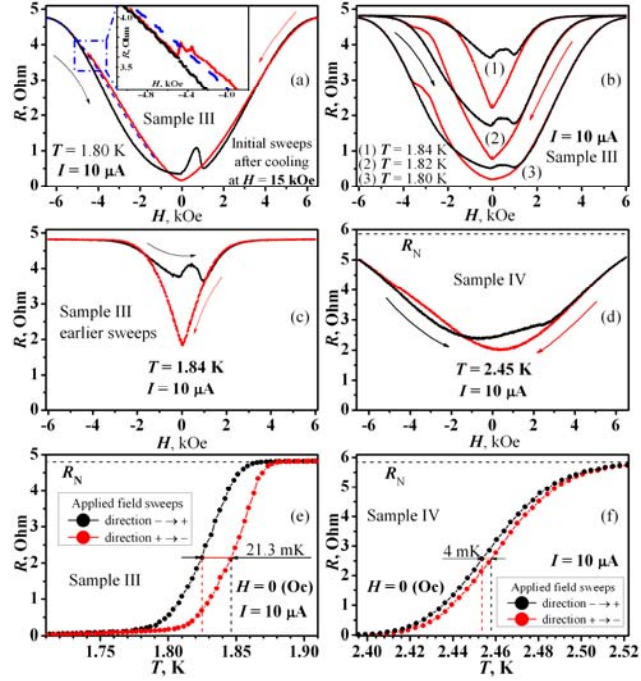

Fig. 3. (a) to (d): $R(H)$ measured at fixed temperatures within the superconducting transition for vanishing field. (e) and (f): Superconducting transitions at $H$=0 measured after different sweeps of the magnetic field. Dashed line in (a): BB sweep $R(H)$ data at positive $H$ reflected across $H$ = 0 to negative $H$.

Upon a further increase of the field magnitude in the negative direction, the torque increases until the exchange biased region of the $Cu_{41}Ni_{59}$ film abruptly reverses at about -4.4 kOe simultaneously with the Co layer magnetic moment (trace the $m(H)$ derivatives in Fig. 2(a)). In this interpretation, the resistance change in Fig. 3(a) at the field of reversal (see the zoomed feature in the insert) is a resistance drop (*i.e.* an increase of $T_c$) due to a vanishing of the triplet component generated by the exchange spring.

The FB of the hysteresis loop, being visually quite smooth, produces a more tricky behavior of $R(H)$ near magnetization reversal, likely because the $Cu_{41}Ni_{59}$ layer (3) passes a multidomain state nearby the reversal (Fig. 2). The upturn feature on $R(H)$ between 0.0 and 1.0 kOe could be treated as a superconducting $T_c$ suppression by the combined action of the triplet pairing generation by noncollinear magnetic configurations in the system[20,21] (yielding a peak of resistance near the perpendicular alignment of magnetizations), and the mechanism of correlated magnetic microstructures, proposed and elaborated in Ref. [19] for F/S/F spin-valve cores. Note here that due to the training effect[24,27], the reversal fields for the Co and bottom $Cu_{41}Ni_{59}$ layers are already split (see Fig. 2(a), especially the derivative of the FB hysteresis loop), and both $Cu_{41}Ni_{59}$ layers are in the multidomain state.

The kink in the resistance at -4.4 kOe of the BB branch could possibly also be interpreted as the vanishing of a stray field induced increase of $R(H)$ due to an inhomogeneous magnetization indicated by the only nearly flat (inclination increasing in trained state) $m(H)$ range between -1.2 and -4 kOe[17-19,37]. Magnetostatic coupling between magnetic microstructures enhances the alignment[38] and intensifies the stray field penetrating the superconductor[19]. Also domain wall (DW) induced superconductivity is possible if the width of the domain wall is small and the saturation magnetization is large[39,40]. However, our previous investigations on $Nb/Cu_{41}Ni_{59}$ layered systems[22,41] indicate that the influence of stray fields of $Cu_{41}Ni_{59}$ layers alone are too weak to result in peak structures.

The $R(H)$ dependence for the trained state after ten full sweeps is given in Fig. 3(b) for three temperatures. The coercive fields $H_c^{BB}$ and $H_c^{FB}$ are split for both branches (as already in the second hysteresis loop, see paired peaks of the derivative in Fig. 2(b)), what results in basically similar, but a bit broadened $R(H)$ features, described for the first loop, and the same physics behind. There is a relatively *weak* magnetic noncollinearity within the bottom $Cu_{41}Ni_{59}$ layer, occurring in the range from about 0 to -2.5 kOe and -4 to -0.5 kOe for the BB and FB, respectively, expressed by a small angle between magnetization arrows in the schematic shown in Fig.2(b) for layers (2) and (3), which is probably responsible for a suppression of the direct SSV in the BB. At the borders of these ranges there are field regions of *strong* noncollinearity within the bottom $Cu_{41}Ni_{59}$ layer, inherent to the reversal regions of BB and FB branches, generating the upturn features in the $R(H)$ curves. The significant difference is that the resistance at zero magnetic field strongly depends on the pre-history of that state (see Fig. 3(c), representing an earlier cycle at $T$ = 1.84 K, at which also the measurement (1) in Fig. 3(b) has been performed). Indeed, around $H$ = 0 in the FB (and $H$ = -3.5 kOe in the BB) the magnetization directions of all three layers are *strongly* noncollinear (see Fig. 2(b)), and, especially important, those of layers (3) and (2) rotate against each other for increasing (and, respectively, decreasing) field, achieving nearly perpendicular alignment. Since layer (3) contains a certain part $x$ of the bottom layer, strongly noncollinear magnetization configurations occur inside this layer and, thus, a strong triplet component generation is expected[20]. The triplet component suppresses $T_c$ leading to an increase of the magnetoresistance[22]. The difference in magnetoresistances in Fig. 3(c) corresponds to a difference in the superconducting $T_c$ of 21 mK, measured directly from the resistive transition as shown in Fig. 3(e), at zero field.

To check if the effect arises from stray fields of the cobalt layer (here, a possible DW width is similar as in $Cu_{41}Ni_{59}$ layers[42,43], but the higher saturation magnetization favors cobalt to produce stray field peaks on magnetoresistance[39,43]), we measured Sample IV (shown in Fig. 3(d)), for which the thickness of the copper-nickel layers is very small (below 1 nm) or vanishing. In accordance with the random anisotropy model[44] the coercivity of $Cu_{41}Ni_{59}$ layers reduces[45], and the coercive loop becomes narrow. So the stray field effect, which is maximal at the coercivity fields, shrinks to a narrow interval of fields around $H$=0. However, the $R(H)$ splitting of the BB and FB extends over 6 kOe, *i.e.* it has another origin. It correlates with the Co hysteresis loop. We attribute this to the stray fields arising from disordered spins at the $Co/CoO_x$ and $CoO_x/Cu_{41}Ni_{59}$ interface in the trained state.[24] The Co layer of 4 nm is expected not to produce substantial stray fields because of Neel type domain wall structure at that film thickness. The triplet pairing generation for such extremely thin copper-nickel layers is also expected to be negligible.[20] Thus, the features in R(H) shown in Fig. 3(d) represent the background contribution of the Co/CoO$_x$ stray fields. The difference in $T_c$ measured at H=0 for the BB and FB is about 4 mK (see Fig. 3(f)).

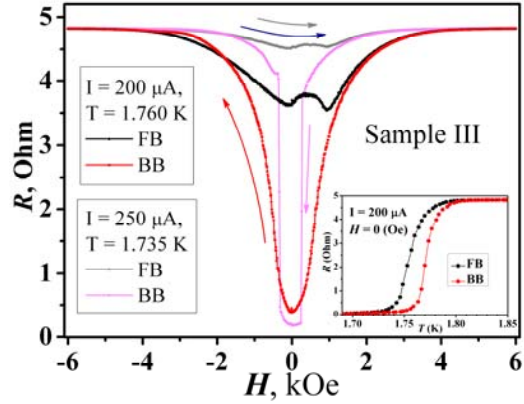

Fig. 4. Magnetoresistance of Sample III at higher measurement currents. Insert: superconducting transitions at zero applied field for FB and BB for the bias current 200 µA.

To check the transport properties of the device structure we measured R(H) for currents in the range 1 - 250 µA. Up to about 200 µA no significant changes were observed. At 200 µA and above a drastic increase of the difference $R_{FB}$-$R_{BB}$ at zero field was registered for Sample III (see Fig. 4). While the resistance $R_{FB}$ approaches the normal state value $R_N$ in the FB, the resistance $R_{FB}$ nearly vanishes, in the BB, resulting in a ratio ($R_{FB}$-$R_{BB}$)/$R_N$ of about 90%. The reason is a sharpening of the superconducting transition, with the width $\delta T_c = T(0.9\ R_N)$-$T(0.1\ R_N)$ decreasing from $\delta T_c$ = 50 mK to 30 mK for 10 µA and 200 µA, respectively (see Figs. 3(e) and 4(insert), $T_c$ is lowered by 70 mK). A plausible explanation could be an avalanche like flux flow at high Lorentz force, due to the increased current[46], or an instability of the flux line motion in the resistive mixed state, caused by an escape of nonequilibrium quasiparticles from the vortex cores[47,48]. This effect has never been observed in Sample IV (vanishing Cu$_{41}$Ni$_{59}$ layers thickness) for any current up to the critical one for any measurement temperature, although the expected critical velocity for the instability is expected to be smaller for Nb than for Nb/Cu$_{41}$Ni$_{59}$[48,49]. This suggests, that the triplet pairing, generated by noncollinear magnetizations in the Cu$_{41}$Ni$_{59}$ layers, is involved in the phenomenon.

In summary, a strong exchange biasing of about 2 kOe for a diluted ferromagnetic copper-nickel alloy was obtained at the interface with a Co/CoO$_x$ bilayer. Combined with an F/S/F SSV core, the Co/CoO$_x$/Cu$_{41}$Ni$_{59}$/Nb/Cu$_{41}$Ni$_{59}$ spin valve exhibits a magnetic memory effect, which depends on the preceding field polarity. High and low resistance states do not require bias fields or currents to keep them in an idle mode. While the writing of a state requires only field sweeps up to magnetic saturation, for reading only a current at H = 0 is necessary. We propose to ascribe the difference in resistances to the training effect and the generation of an odd in frequency triplet pairing component at noncollinear alignment of the magnetizations in the system around H = 0. Moreover, e.g. stray field effects of the micromagnetic structures of the layers may contribute to the effect. Since the mechanisms described above provide different critical temperatures for the two logical states, one does not need to realize the hardly achievable antiparallel magnetic alignment required for the direct SSV effect.

**Acknowledgments**

The authors are grateful to S. Heidemeyer, B. Knoblich and W. Reiber for assistance in the TEM sample preparation, and to D. Vieweg for assistance in magnetic measurements. The work was supported by the Deutsche Forschungsgemeinschaft (DFG) under the grant No GZ: HO 955/6-2. In part the Russian Fund for Basic Research (RFBR) supported the project under the grant 11-02-00848-a (LRT). The magnetic investigations (H.-A. K.v.N.) were partially supported by the Deutsche Forschungsgemeinschaft (DFG) within the Transregional Collaborative Research Center TRR 80 "From Electronic Correlations to Functionality" (Augsburg, Munich).

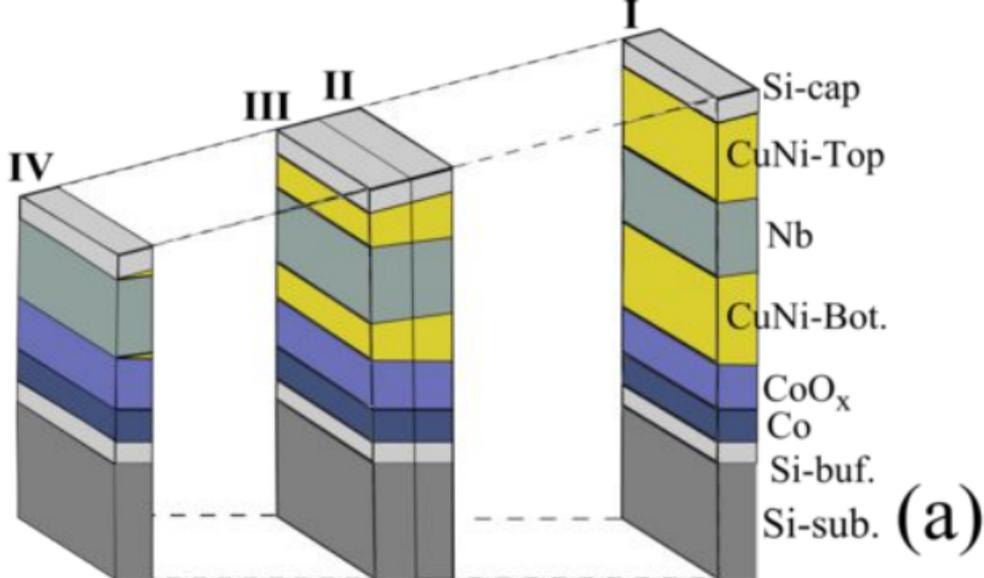
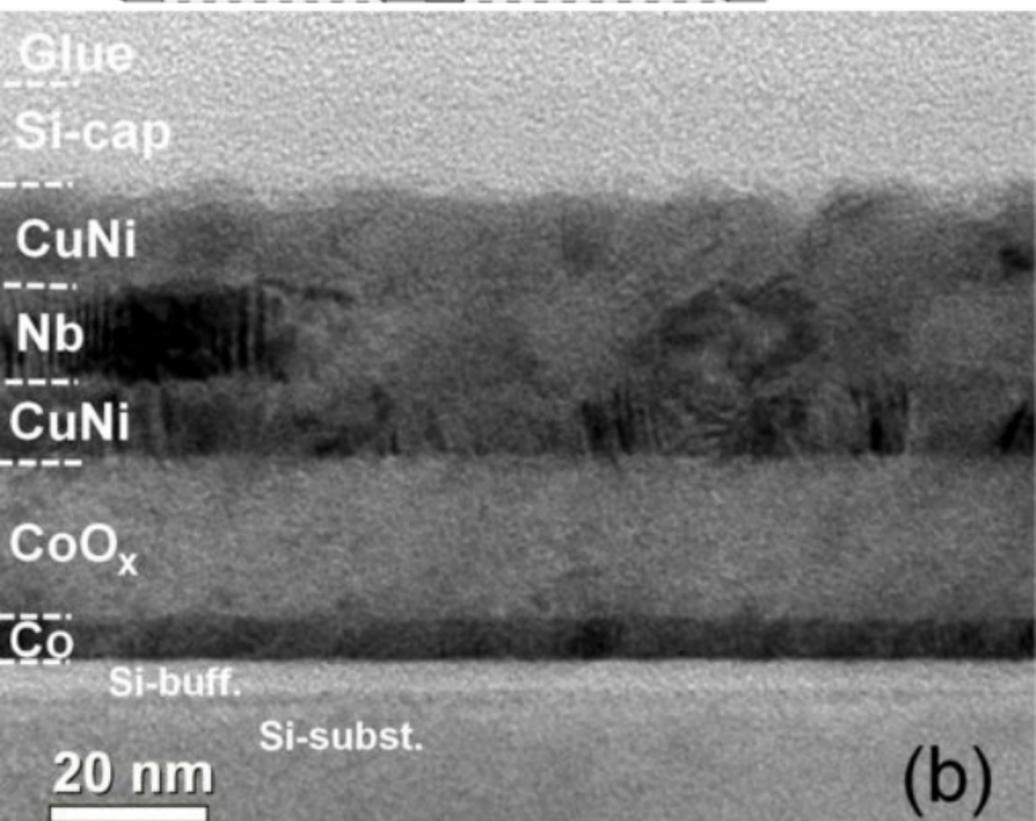

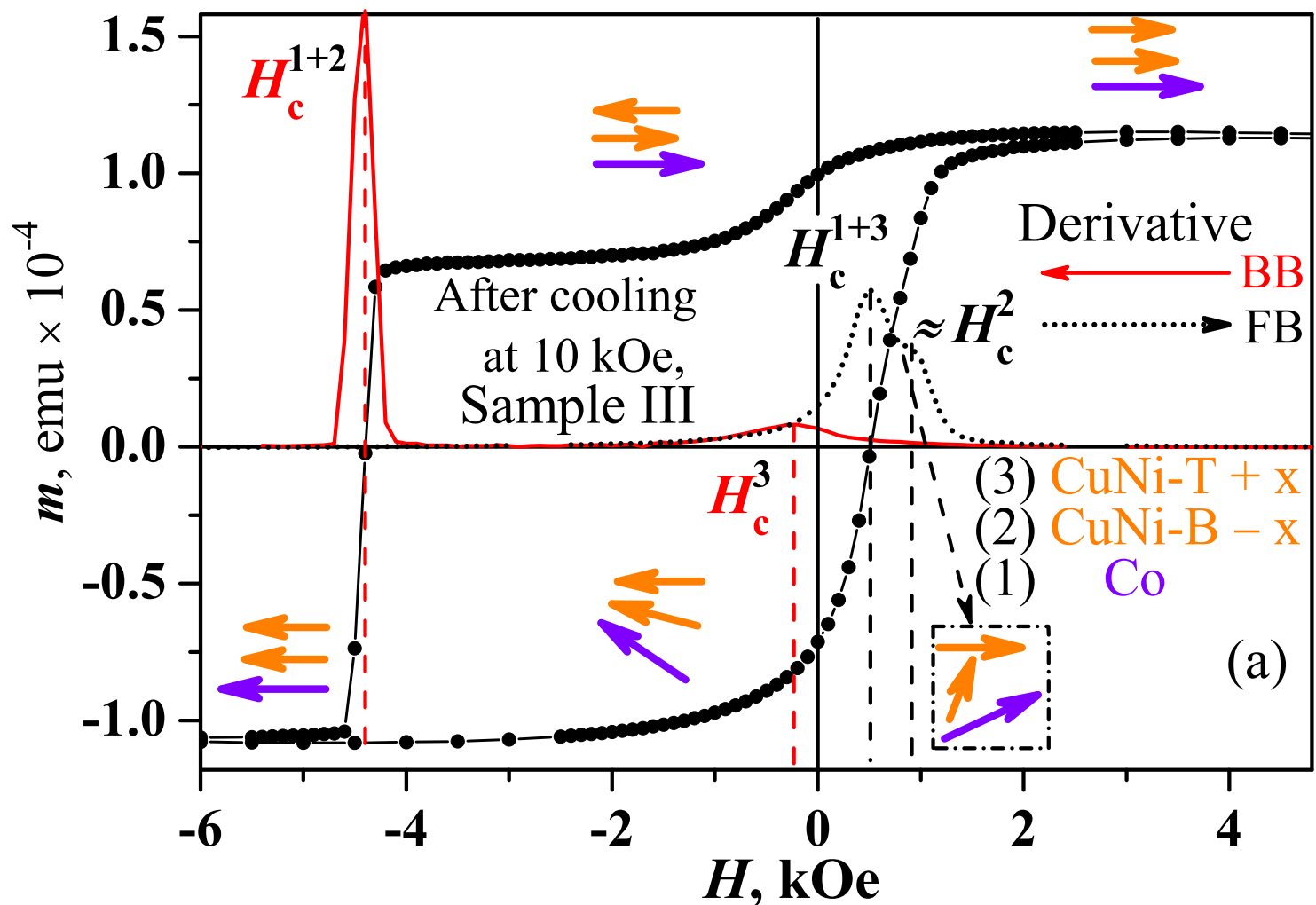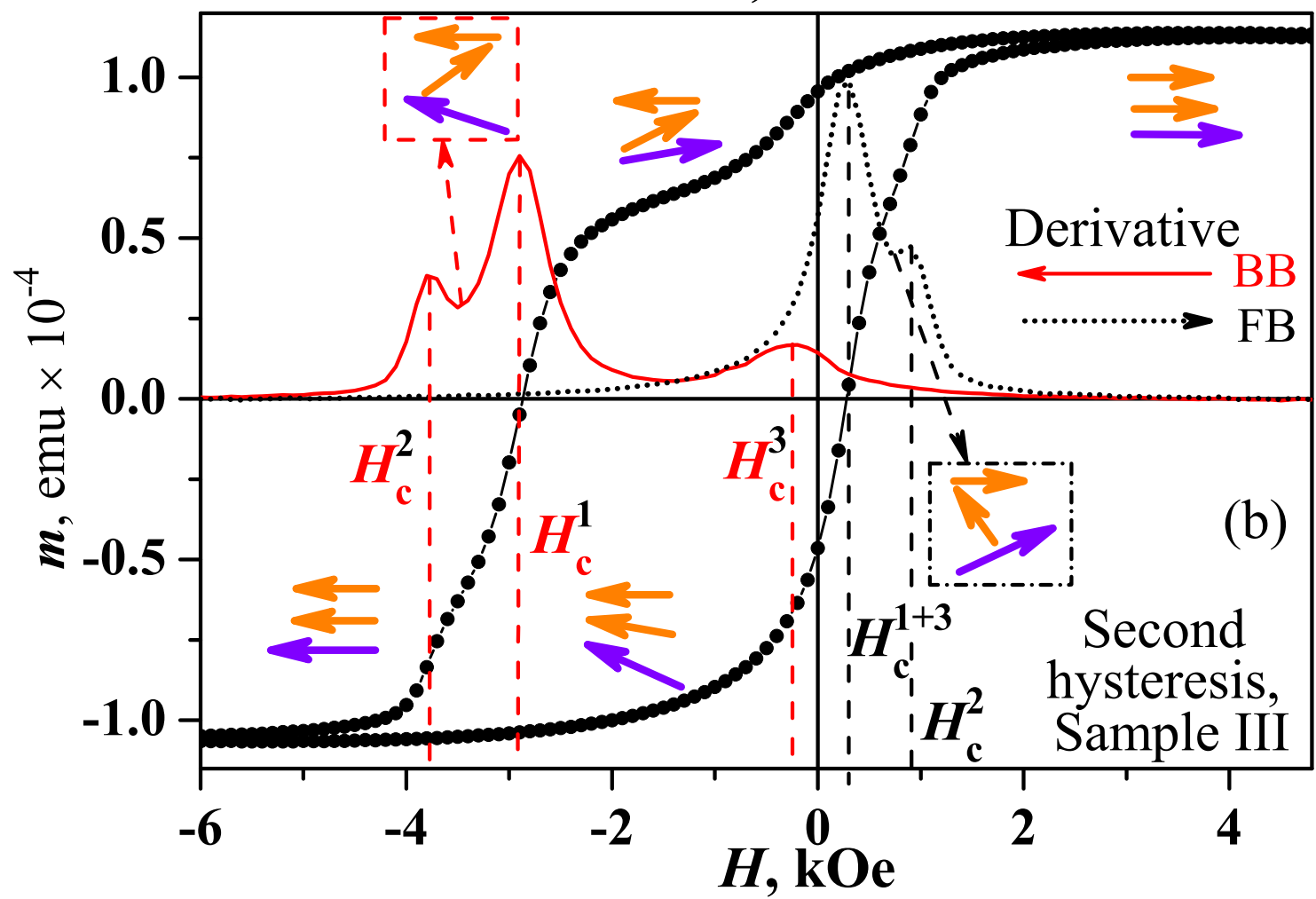

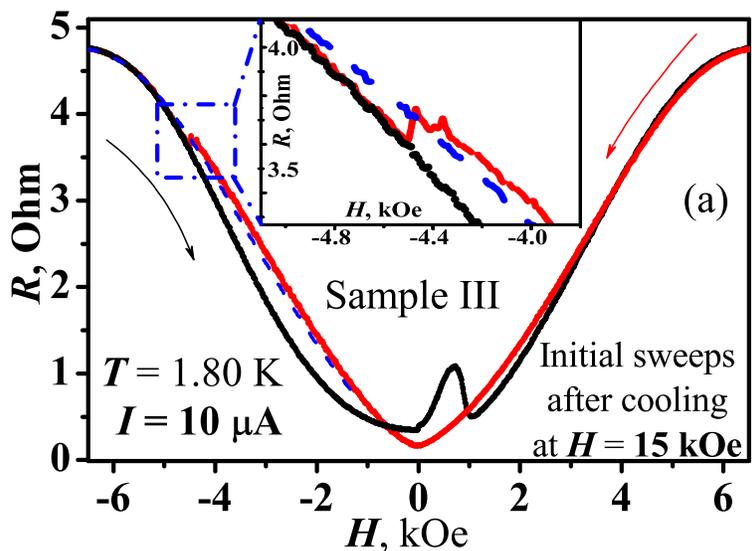
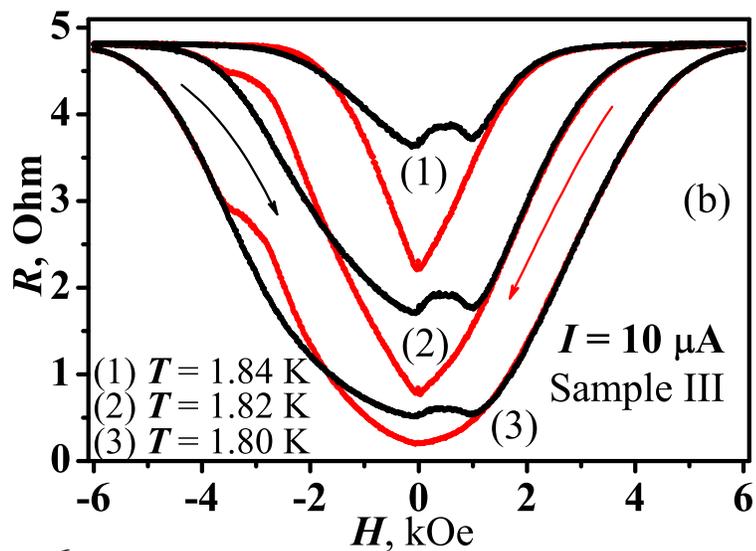
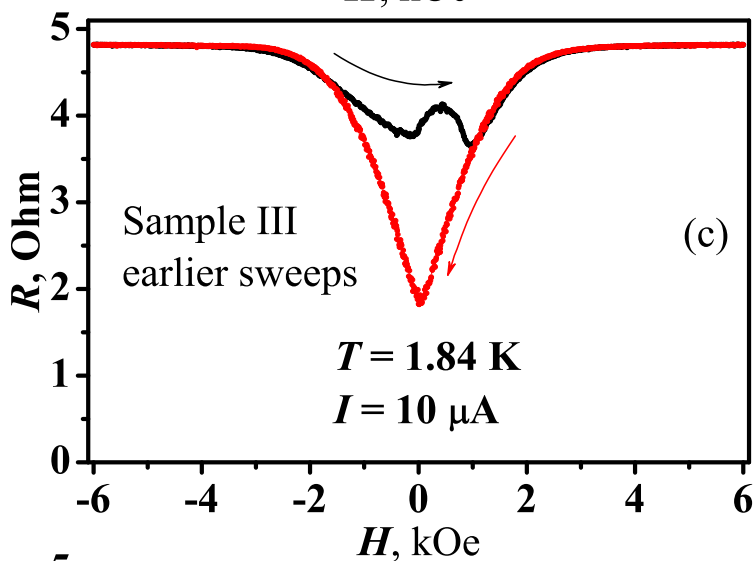
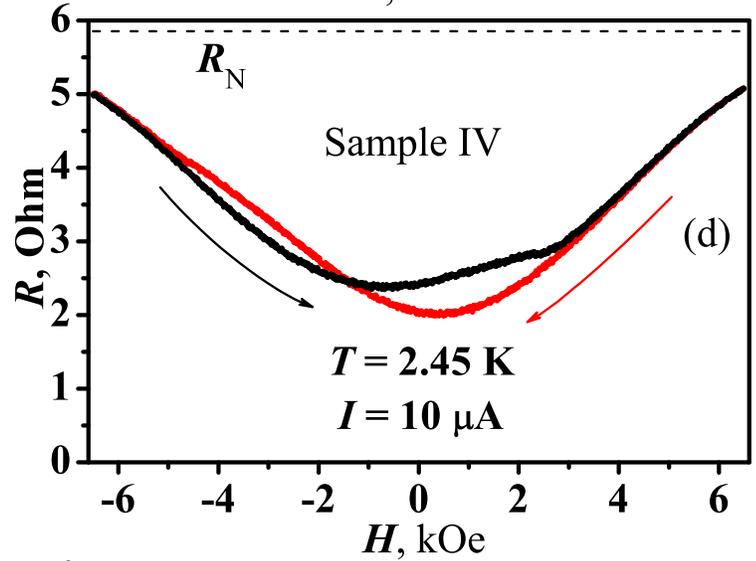
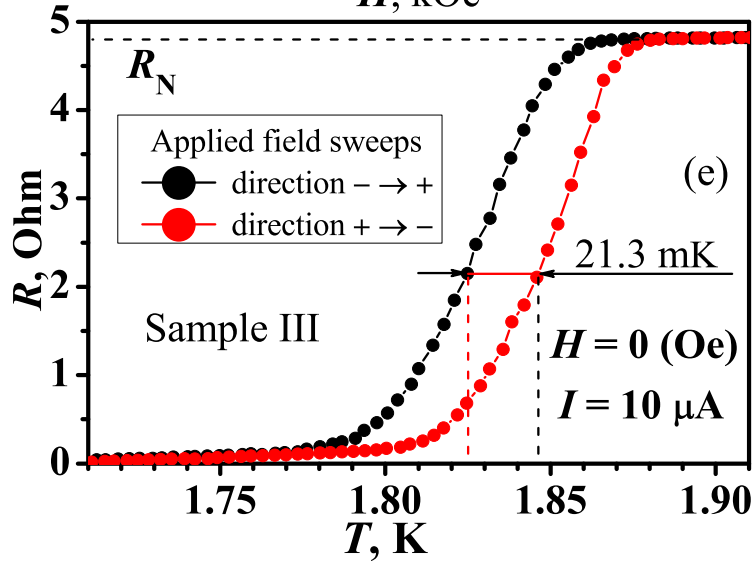
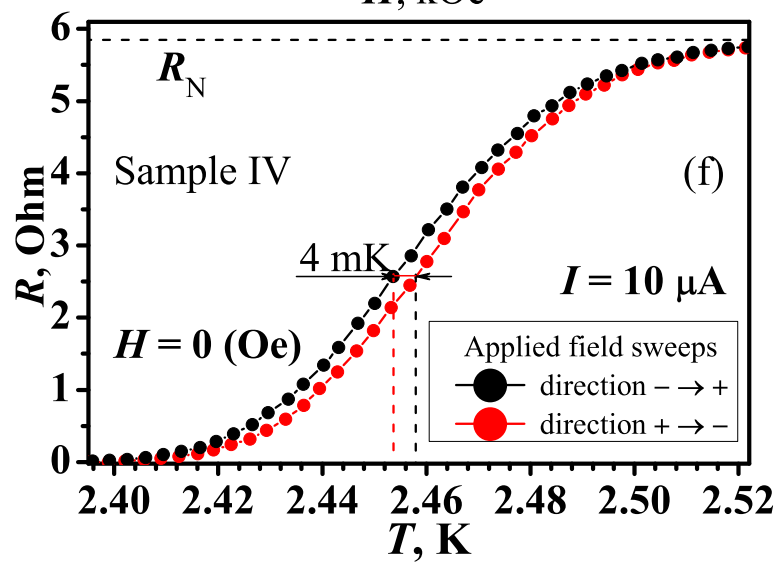

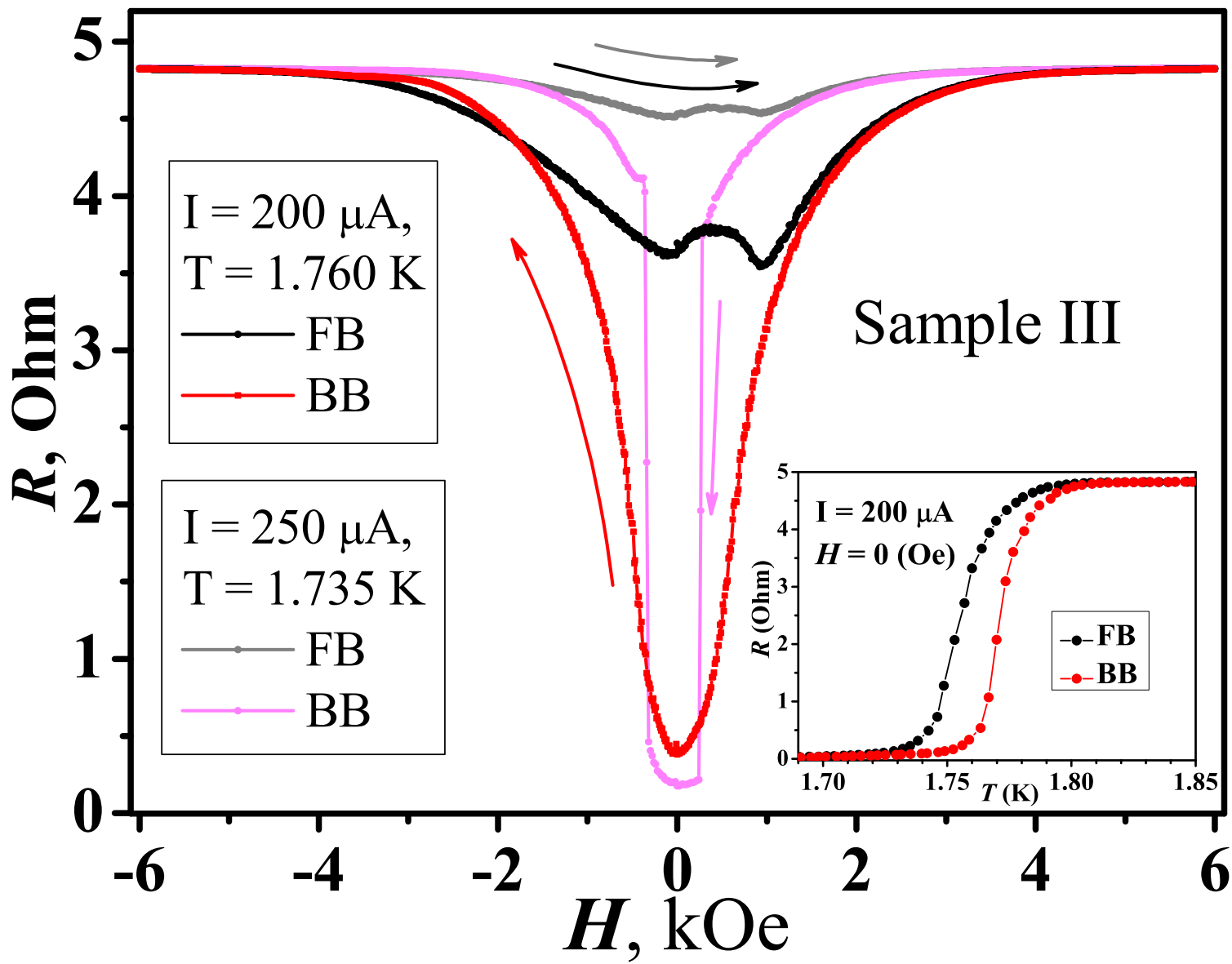